https://doi.org/10.3847/1538-4357/ad09b5

# Surface Volatile Composition as Evidence for Hydrothermal Processes Lasting Longer in Triton's Interior than Pluto's

Kathleen Mandt[1] , Adrienn Luspay-Kuti[2] , Olivier Mousis[3] , and Sarah E. Anderson[3]
[1] NASA Goddard Space Flight Center, Greenbelt, MD 20771, USA; Kathleen.Mandt@nasa.gov
[2] Johns Hopkins University Applied Physics Laboratory, 11100 Johns Hopkins Rd., Laurel, MD 20723, USA
[3] Aix Marseille Université, Institut Origines, CNRS, CNES, LAM, Marseille, France

## Abstract

Ocean worlds, or icy bodies in the outer solar system that have or once had subsurface liquid water oceans, are among the most compelling topics of astrobiology. Typically, confirming the existence of a subsurface ocean requires close spacecraft observations. However, combining our understanding of the chemistry that takes place in a subsurface ocean with our knowledge of the building blocks that formed potential ocean worlds provides an opportunity to identify tracers of endogenic activity in the surface volatiles of Pluto and Triton. We show here that the current composition of the volatiles on the surfaces and in the atmospheres of Pluto and Triton are deficient in carbon, which can only be explained by the loss of $CH_4$ through a combination of aqueous chemistry and atmospheric processes. Furthermore, we find that the relative nitrogen and water abundances are within the range observed in building block analogs, comets, and chondrites. A lower limit for N/Ar in Pluto's atmosphere also suggests source building blocks that have a cometary or chondritic composition, all pointing to an origin for their nitrogen as $NH_3$ or organics. Triton's lower abundance of $CH_4$ compared to Pluto, and the detection of $CO_2$ at Triton but not at Pluto points to aqueous chemistry in a subsurface ocean that was more efficient at Triton than Pluto. These results have applications to other large Kuiper Belt objects as well as the assessment of formation locations and times for the four giant planets given future probe measurements of noble gas abundances and isotope ratios.

*Unified Astronomy Thesaurus concepts:* Solar system formation (1530); Planet formation (1241); Pluto (1267); Triton (2187)

## 1. Introduction

Triton and Pluto are of great interest because of their similar origins yet very different evolutionary histories. By studying both we can gain insight into the role of volatiles in the outer solar system, both in forming the giant planets and small bodies found in the Kuiper Belt objects (KBOs) as well as determining the conditions required to form subsurface liquid water oceans.

We seek to understand the formation conditions of Pluto and Triton using the composition of their building blocks to determine where they formed. This information tells us about the conditions of the protosolar nebula (PSN) during their formation and has implications for understanding the formation regions of the giant planets. Of particular interest for Pluto and Triton is the origin of nitrogen and carbon that is observed in their atmospheres and on their surfaces. To constrain their origin, we evaluate the range of possible compositions of the building blocks that formed Pluto and Triton based on knowledge of both their current compositions as well as processes that can change the compositions.

The current composition of Pluto and Triton provides the starting point for determining their origins because the tracers of their building blocks are found on their surfaces and in their atmospheres today. Although the current state of knowledge is limited, there is information presently available about the bulk density of each body and the composition of their atmospheres and surface ices. Furthermore, upper limits available for argon

(Ar) and $HC^{15}N$ in Pluto's atmosphere provide tantalizing clues to the origin of nitrogen (N) and carbon (C) and the geochemical activity levels in the interior of each.

The processes that modified the composition of Pluto and Triton include chemistry within the interior after differentiation and formation of a liquid water ocean, formation of layers of clathrates within the ocean, atmospheric chemistry, and escape from the atmosphere to space. Many of these processes will change the molecular composition but will not change the bulk composition of elements. For example, the conversion of $NH_3$ into $N_2$ and $CH_4$ to $CO_2$ in the interior will change the bulk molecular composition, but the bulk nitrogen-to-carbon ratio, or N/C, will remain the same. To simplify the evaluation of Pluto's and Triton's composition, we will evaluate the bulk elemental ratios N/C, O/C, N/O, O/H, and N/Ar.

### 1.1. Origin of Nitrogen

Although we know that the building blocks of Pluto and Triton formed in the outer solar system, it is not clear based on our current understanding of solar system formation if the temperature and pressure conditions allowed them to trap sufficient nitrogen in the form of $N_2$ to provide the $N_2$ observed on their surfaces and in their atmospheres. $N_2$ requires temperatures below ∼40 K to be trapped in either amorphous (Bar-Nun et al. 1985, 1988) or crystalline (Mousis et al. 2012, 2014) water ice. Alternatively, the nitrogen observed today could have originally been trapped as $NH_3$ ice or organics and later converted to $N_2$, as has been found for Titan (Mandt et al. 2014; Miller et al. 2019).

In the PSN, $N_2$ is proposed to have been 10 times more abundant than $NH_3$ (Lewis & Prinn 1980), but $NH_3$ is preferentially incorporated into building blocks due to its ability to condense or be trapped in solid ice at higher temperatures than $N_2$. Conversion of $NH_3$ to $N_2$ could happen through atmospheric chemistry (Atreya et al. 1978), aqueous chemistry in the interior (Glein et al. 2008), or through impact processes (McKay et al. 1988; Ishimaru et al. 2011; Sekine et al. 2011). Organics in the PSN are thought to have an abundance of nitrogen relative to carbon (N/C) of ~0.04 (Kissel & Krueger 1987; Alexander et al. 2017; Fray et al. 2017). This means that an organic origin for nitrogen is possible, but would require *cooking* of the organics at very high temperatures in the interior (Miller et al. 2019; McKinnon et al. 2021) that would also produce large abundances of carbon-bearing species.

At the present time, comets are the best analog for the building blocks of Pluto and Triton because they may have formed in similar conditions. It is important to note that comets are a combination of ice that forms a coma and refractory materials that are thought to be best represented by chondrites (Joswiak et al. 2017; Mandt et al. 2022). Therefore, when discussing cometary analogs for the building blocks of Pluto and Triton we refer to a combination of cometary ices and chondritic materials. The Rosetta spacecraft provided the first detailed measurements of nitrogen-bearing molecules in the coma of a comet providing a ratio of $N_2$ to $NH_3$ of $0.13 \pm 0.05$ in the ices of 67P/Churyumov–Gerasimenko (67P/C-G; Rubin et al. 2019). This ratio, and the lack of detection of $N_2$ in most comets (e.g., Cochran 2002; Anderson et al. 2023), suggests that they are deficient in $N_2$ relative to $NH_3$. This deficiency would either result from formation temperatures too high to trap significant amounts of $N_2$ (Iro et al. 2003) or preferential loss of $N_2$ resulting from internal radiogenic heating at early epochs after formation (Mousis et al. 2012) and additionally during their first pass through the solar system (Owen et al. 1993). The detection of $N_2$ in 67P/C-G is important because it showed that $N_2$ is present in short period comets thought to originate in the same region where Pluto and Triton formed (Rubin et al. 2019).

Another class of comets may have also formed when the solar system formed as indicated by the composition of C/2016 R2 PanSTARRS, which is heavily depleted in $H_2O$ and has an unusually high $N_2$/CO abundance ratio based on observations of ions in the coma (McKay et al. 2019). No $NH_3$ abundances are available for this comet, limiting the use of its composition for elemental analyses to lower limits for N/O and N/C. This comet was initially proposed to be the fragment of a differentiated object, similar to current-day Pluto (Biver et al. 2018), but preserving blocks of $N_2$ ice during the massive collisions required to break apart such an object is difficult to explain (Bergner & Seligman 2023). A more realistic proposed origin is that R2 formed near the CO and $N_2$ ice line (Mousis et al. 2021; Price et al. 2021) and represents a different class of comets than the ones noted to be $N_2$ depleted that were mostly ejected from the solar system during giant planet migration (Anderson et al. 2022).

A recent study comparing the $N_2$/$H_2O$ ratio of 67P/C-G suggested that Pluto's $N_2$ could be primordial (Glein & Waite 2018). However, this study noted that the $N_2$/CO ratio at Pluto could only agree with a cometary origin if large amounts of CO had been removed through a variety of processes (Glein & Waite 2018; McKinnon et al. 2021). Furthermore, the abundance of $NH_3$ in 67P/C-G is much greater than the abundance of $N_2$ resulting in a much higher total abundance of nitrogen relative to water than suggested by the $N_2$/$H_2O$ ratio. This means that Pluto, and Triton by extension based on their similarities, could have obtained a larger abundance of the observed nitrogen in its primordial ices in the form of $NH_3$, similar to what happened at Titan (Mandt et al. 2014).

### 1.2. Origin of Carbon

Previous studies have focused primarily on nitrogen to trace the formation conditions of Pluto (e.g., Mandt et al. 2017; Glein & Waite 2018; McKinnon et al. 2021). However, the origin of carbon-bearing volatiles, $CH_4$ and CO at Pluto (Protopapa et al. 2017) and Triton (Cruikshank et al. 1984, 1993), as well as the detection of $CO_2$ at Triton (Cruikshank et al. 1993) but not Pluto (Grundy et al. 2016) can also provide valuable constraints.

Of particular interest is the influence of aqueous chemistry and atmospheric processes on the composition and abundance of carbon-bearing volatiles. Aqueous chemistry would have taken place after the formation of Pluto and Triton, when differentiation of the interior and heating may have produced a subsurface ocean where hydrothermal processes changed the composition of the molecular species (Shock & McKinnon 1993; Glein & Waite 2018). These processes lead to conversions between $NH_3$ and $N_2$ and between $CH_4$ and $CO_2$. Depending on the internal temperature and pressure, the pH in the water, the bulk concentration of N, and the oxidation state of the system, the conditions can either favor the production of $N_2$ and $CO_2$ or $NH_3$ and $CH_4$. Formation of $N_2$ and $CO_2$ is favored at higher temperatures, lower pressures, and with systems that are more oxidized (Glein et al. 2008). Ultimately, aqueous chemistry will change the molecular composition of the volatiles that are present but will not affect the bulk elemental composition, such as N/C.

Additional aqueous reactions have been identified that can lead to permanent loss of CO (Shock & McKinnon 1993; Neveu et al. 2015; Glein & Waite 2018). This would change bulk elemental ratios like N/C from what existed in the building blocks to values with reduced amounts of carbon and oxygen, but would not influence C/O significantly.

We outline in Section 2 what is known about the compositions of Pluto and Triton. We discuss the results in Section 3 and processes that will fractionate, or alter, the elemental ratios to match the observations. In Section 4, we summarize the implications for the composition of the building blocks of Pluto and Triton.

## 2. The Composition of Pluto and Triton

Triton and Pluto have been of great interest to the planetary science community because of their similar sizes and densities but very different histories. Pluto is the largest KBO. Its orbit is tilted off of the ecliptic by 17° and has an eccentricity of 0.25. A year on Pluto lasts 248 Earth yr and brings Pluto inside of Neptune's orbit for part of its year. It is possible that Pluto formed closer to the Sun, and the migration of Neptune placed it in its current orbit (see McKinnon et al. 2021 and references therein). Triton, on the other hand, is the largest moon of Neptune and is thought to have formed in the same

**Table 1**
Estimated Moles of $N_2$ and $H_2O$ in the Current Bulk Composition of Pluto and Triton Based on the Bulk Density for Each

|  | Moles of $H_2O$ | Moles of $N_2$ |
| --- | --- | --- |
| Pluto | $(2-2.58) \times 10^{23}$ | $(0.4-3) \times 10^{20}$ |
| Triton | $(1.5-2.03) \times 10^{23}$ | $(0.1-1.4) \times 10^{21}$ |

**Note.** The surface inventory of nitrogen is assumed to be representative of the bulk inventory.

region as Pluto but had been captured by Neptune (Agnor & Hamilton 2006).

To constrain the current composition of Pluto and Triton, we will start with the available constraints for the bulk water abundance based on the bulk density of each of them. We will then evaluate the nitrogen and carbon abundances based on observations of the surface and atmosphere. We will end with a discussion of the only available constraints on noble gases and isotopes, which are upper limits for Ar and HC$^{15}$N in Pluto's atmosphere.

### 2.1. Bulk Water Abundance

$H_2O$ ice has been detected on the surfaces of both Pluto and Triton and is believed to be the main component in their bedrock (Cruikshank et al. 1993; Grundy et al. 2016; Dalle Ore et al. 2018, 2019; Cook et al. 2019). The density of Pluto and Triton provides some potential information about the fractions of rock and water present, both in the form of ice and potentially as a subsurface liquid ocean. The average bulk density of Pluto and Triton are 1854 kg m$^{-3}$ (Nimmo et al. 2017) and 2061 kg m$^{-3}$ (McKinnon & Kirk 2014), respectively. Using this and the densities of water and rock, Glein & Waite (2018) estimated the mass fraction of the water for Pluto to be between 0.28 and 0.36. Using their methods and values for rock and water density, we determine a range of 0.21–0.28 for water at Triton. Next, using the masses of Pluto and Triton and converting to moles, we can determine the bulk abundance of $H_2O$ to be $(2-2.58) \times 10^{23}$ mol and $(1.5-2.03) \times 10^{23}$ mol, respectively.

### 2.2. Nitrogen

Infrared absorption measurements show that Pluto's surface $N_2$ reservoir is mainly concentrated to Sputnik Planitia (Grundy et al. 2016; Protopapa et al. 2017). The surface reservoir is estimated to outweigh the atmospheric $N_2$ reservoir by orders of magnitude (Table 1; Glein & Waite 2018), making it a reasonable first-order assumption to use this to represent the current amount of bulk nitrogen for Pluto (Glein & Waite 2018; McKinnon et al. 2021). The apparent amount of nitrogen in Sputnik Planitia is $(0.4-3) \times 10^{20}$ mol (Glein & Waite 2018).

Although no Voyager 2 absorption measurements are available for Triton's surface, we can estimate the volatile abundances by using the Voyager 2 atmospheric measurements, the energy-limited mass flux of $N_2$, and limits to the thickness of the $N_2$ polar deposits, estimated to be no more than 0.5–1 km thick averaged over the surface of Triton (McKinnon et al. 1995). Using these values, the amount of nitrogen on Triton is estimated to be $(0.7-1.4) \times 10^{21}$ mol (see Table 1). We note that these values give an uncertainty of only a factor of 2 compared to Pluto, which has more extensive observations yet an uncertainty that is much larger. To address the limited amount of data available for Triton, we assume a value of $(0.1-1.4) \times 10^{21}$ mol for Triton.

### 2.3. Carbon

The carbon abundance can be estimated based on the abundances of $CH_4$, CO, and $CO_2$ in the atmosphere and on the surface. The atmospheres of both Pluto and Triton are predominantly $N_2$ with minor amounts of $CH_4$ and CO. Triton and Pluto have only been explored by a single spacecraft flyby of each. While the New Horizons flyby of Pluto provided great detail on the surface and atmospheric composition of Pluto, the Voyager 2 flyby only provided the atmospheric composition of Triton because the spacecraft lacked the instrumentation needed to study the surface composition. As such, the surface composition of Triton is determined using ground-based observations.

Pluto's neutral atmosphere composition was determined during the New Horizons flyby to be >99% $N_2$, ~0.30% $CH_4$ (Young et al. 2018), and ~0.05% CO (Lellouch et al. 2017). Trace amounts of ethane ($C_2H_6$), acetylene ($C_2H_2$), and ethylene ($C_2H_4$) were also detected with atmospheric mixing ratios of 0.001 (or 0.1%) in the middle atmosphere, dropping to mixing ratios between $10^{-7}$ and $10^{-5}$ at ~100 km (Young et al. 2018). The near-surface CO/$N_2$ and the CO/$CH_4$ ratios were $4 \times 10^{-3}$ and $1.7 \times 10^{-3}$, respectively.

The $CH_4$ abundance of Triton's atmosphere was 1 order of magnitude less (~0.01%) during the Voyager 2 flyby. On the other hand, the CO abundance appears to be greater on Triton. The CO/$CH_4$ ratio of ~3.5 is roughly 3 orders of magnitude higher on Triton than on Pluto, primarily because of the lower $CH_4$ abundance, while the CO/$N_2$ ratio of ~0.01 is about a factor of 6 larger at Triton (Cruikshank et al. 1993).

The surface ices of both Pluto and Triton are dominated by $N_2$, with varying amounts of $CH_4$ and CO. Because Voyager 2 was not able to map the surface composition, we have limited knowledge of the distribution of the volatile ices on Triton's surface. We do know that seasonally changing solar insolation is expected to cause the volatile ices to migrate across the surface (Cruikshank et al. 1984, 1993; Bauer et al. 2010; Buratti et al. 2011). As noted earlier, Pluto's $N_2$ ice appears to be concentrated in the Sputnik Planitia basin. The second most abundant component, $CH_4$, is more widely distributed across the surface and has a mole fraction of $(3-3.6) \times 10^{-3}$ diluted in $N_2$ (Protopapa et al. 2017) on Pluto's surface. On Triton's surface, $CH_4$ is predicted to be incorporated in the surface $N_2$ ice at a mole fraction of ~$(1-5) \times 10^{-4}$, which yields a global equivalent layer of ~1 m on Triton's surface (Cruikshank et al. 1993).

New Horizons detected small amounts of CO ice, $H_2O$ ice, and $NH_3$ hydrates on Pluto's surface (Grundy et al. 2016; Dalle Ore et al. 2018, 2019; Cook et al. 2019). The CO mole fraction relative to $N_2$ is $(2.5-5) \times 10^{-3}$ on Pluto's surface (Owen et al. 1993; Merlin 2015). Estimates for Triton's CO/$N_2$ surface mole fraction may vary between <0.01 and 0.15 depending on the atmospheric CO/$N_2$ estimate used, and whether one assumes CO is mixed in the ice, or if it is physically separate from the $N_2$ ice (Cruikshank et al. 1993; Lellouch et al. 2010).

Detectable amounts of $CO_2$ have been observed on Triton's surface (Cruikshank et al. 1993), most likely in the form of exposed ice deposits. No $CO_2$ has been detected on Pluto's surface. The estimated range of possible abundances of the carbon-bearing species relative to $N_2$ are summarized in Table 2.

**Table 2**
Mole Fractions of Carbon-bearing Species Relative to $N_2$ Assuming an Ice-rich Scenario for the Interior of Pluto and Triton, along with Available Constraints for Noble Gas Abundances and Isotopes for Pluto

|  | Pluto | Triton |
| --- | --- | --- |
| $CH_4$ | $(3-3.6) \times 10^{-3}$ | $(1-5) \times 10^{-4}$ |
| CO | $(2.5-5) \times 10^{-3}$ | $<0.01-0.15$ |
| $CO_2$ | Not detected | $(1.5-100) \times 10^{-3}$ |
| Ar | $< 2.16 \times 10^{-4}$ at 400 km | No constraints |
| $^{14}N/^{15}N$ | $>125$ | No constraints |

**Table 3**
Derived Elemental Abundances for Pluto and Triton Based on Their Bulk Densities and the Composition of the Surface Ices and Atmospheres

|  | Pluto | Triton |
| --- | --- | --- |
| N/C | 233–364 | 10–1250 |
| O/C | $(1-8) \times 10^5$ | $(0.7-5) \times 10^4$ |
| N/O | $(0.3-3.0) \times 10^{-3}$ | 0.0014–0.019 |
| O/H | 0.5 | 0.5 |
| N/Ar | $>9.26 \times 10^3$ | n/a |

### 2.4. Noble Gases and Isotopes

Observations of the HCN abundance in Pluto's atmosphere were made by the Atacama Large Millimeter Array (ALMA) providing an upper limit for the column density of $HC^{15}N$ that constrains the $^{14}N/^{15}N$ of HCN in Pluto's atmosphere to be >125 (Lellouch et al. 2017). Unfortunately, this constraint encompasses all possible values for the primordial $^{14}N/^{15}N$ in the building blocks for Pluto and Triton (Mandt et al. 2017; Glein 2023).

Although the noble gas Ar was not detected by New Horizons in Pluto's atmosphere, analysis of the UV spectra provided an upper limit of 6% of the column density of $CH_4$ at ~400 km (Steffl et al. 2020). If atmospheric dynamics can be well constrained, this can be used to determine a lower limit for the N/Ar ratio, which is directly relevant to the formation conditions of the building blocks of Pluto. No Ar abundance or upper limit is available for Triton.

### 3. Results and Discussion

Knowing the current constraints for the volatile composition of Pluto and Triton, we can now estimate the elemental abundances of each. These values are then compared to the bulk composition of the PSN represented as *solar*, to what is known for Jupiter's atmosphere, and to relevant compositions of the potential building blocks in the form of cometary ice and chondrites.

#### 3.1. Elemental Abundances

We use the estimates outlined in Tables 1 and 2 to determine the range of possible values for the current elemental abundances for Pluto and Triton. These results are provided in Table 3.

#### 3.2. Carbon Depletion

The elemental ratios N/C and O/C provide information on the bulk abundance of carbon relative to nitrogen and oxygen in Pluto and Triton compared to the composition of building block analogs. We illustrate in Figure 1 the composition of Pluto and Triton presented in Table 3 compared to the solar composition representing the bulk composition of the PSN (Lodders 2021), Jupiter's composition measured by the Galileo Probe Mass Spectrometer (GPMS; Wong et al. 2004), and the Juno Microwave Radiometer (MWR; Li et al. 2017, 2020), cometary ice from the comet 67P/C-G and an average of estimated ice composition based on coma measurements from several other comets (Rubin et al. 2019), and refractory material from asteroids and comets based on the organics observed in several types of chondrites (Alexander et al. 2017).

What is most notable in this figure is the extreme depletion of carbon compared to nitrogen and oxygen in the observed volatiles of Pluto and Triton. Both have supersolar, or greater than the solar values, N/C and O/C by 1.5–4 and 3–6 orders of magnitude, respectively. Compared to the composition of the analogs for the building blocks, which have subsolar to slightly subsolar N/C and supersolar O/C, Pluto and Triton also have a depletion in carbon of 2–4 orders of magnitude. Although the N/C of comet C/2016 R2 is only a lower limit, the comparison with Pluto and Triton still shows it to be rich in carbon compared to their known volatiles, a point initially recognized by Stern et al. (1997). Because no cometary or chondritic analogs for building blocks have been found with ratios like what is observed using the surface and atmosphere for Pluto and Triton, we rule out the primordial building block composition as the origin of this carbon depletion for Pluto and Triton. Therefore, some process must have occurred to remove carbon relative to nitrogen and oxygen.

Several processes were identified in Section 1 that could change the volatile abundances for Pluto and Triton. Previous work has shown how CO can be preferentially removed to explain the very high $N_2$/CO ratio of Pluto compared to comets (Glein & Waite 2018; McKinnon et al. 2021). However, the removal of CO is limited in how much it can increase the O/C ratio. Using the average cometary abundances of CO and $CH_4$ to estimate the change in O/C by removing all of the CO, we find that O/C increases from ~15 to ~135, which is less than 1 order of magnitude. As observed here, O/C is enhanced by 2–6 orders of magnitude requiring significantly more removal of carbon. Therefore, processes that remove CO will not resolve the extreme carbon depletion observed at Pluto and Triton.

Similarly, removing $CO_2$ faces the same problem with the O/C ratio. The best explanation for the carbon depletion would be the removal of carbon through the loss of primordial $CH_4$. There are two ways to preferentially remove $CH_4$ compared to species containing nitrogen and oxygen: conversion to $CO_2$ in the interior through aqueous chemistry (Glein et al. 2008), and atmospheric processes that include photochemical conversion to more complex organics combined with escape from the top of the atmosphere (Mandt et al. 2009, 2012).

Removal of $CH_4$ by conversion to $CO_2$ in the interior would produce large amounts of $CO_2$. As noted in Section 1, models for aqueous chemistry show that the production of $CO_2$ is thermodynamically preferred at higher temperatures, lower pressures, and in more oxidized systems with higher pH (Glein et al. 2008). The conversion of primordial $NH_3$ to $N_2$ is also favored by the same conditions. Although no $CO_2$ has been detected at Pluto, of all species discussed here $CO_2$ is the most stable clathrate guest, or molecule that can be trapped in water ice cages (Sloan & Koh 2007), so $CO_2$ produced through this process could be trapped in clathrates in a subsurface ocean (Kamata et al. 2019). Triton on the other hand appears to have more $CO_2$ than $CH_4$, and less $CH_4$ than Pluto. Furthermore, $NH_3$ has been observed on the surface of Pluto in localized occurrences possibly related to recent geologic activity (Dalle Ore et al. 2019) but has not been reported at Triton. Although the lack of detection may be due to the difficulty of resolving $NH_3$ bands on a methane-covered surface from hemisphere-scale remote sensing, these combined observations suggest that interior conversion of $NH_3$ and $CH_4$ to $N_2$ and $CO_2$ may have been a more complete process at Triton than at Pluto.

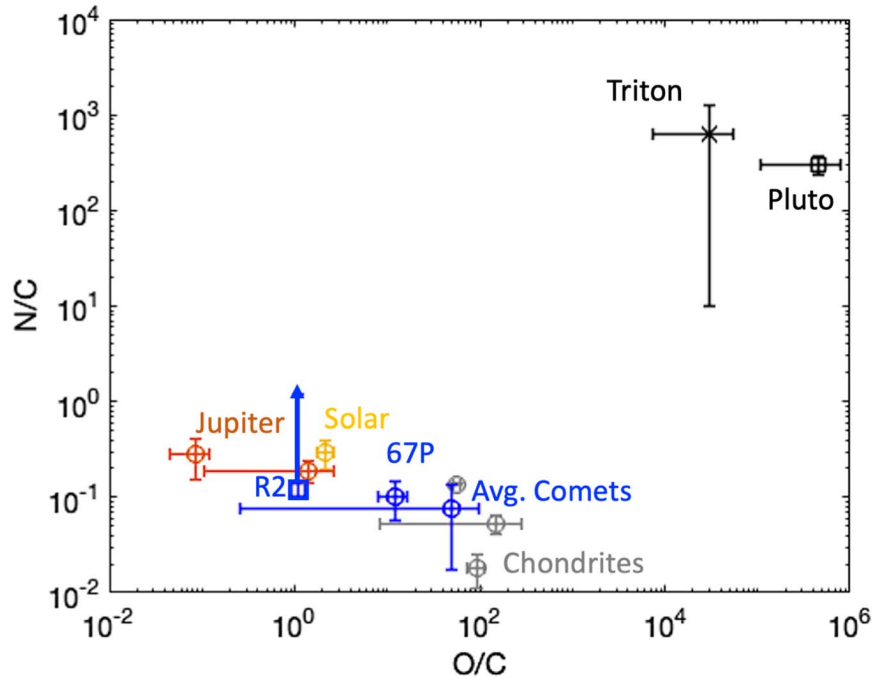

**Figure 1.** Elemental ratios providing information on the relative carbon abundance observed for Pluto and Triton. These ratios are compared to the solar abundances, abundances observed in Jupiter's atmosphere, and analogs for solid materials in the PSN. The two measurements for Jupiter were made by the Galileo Probe Mass Spectrometer (GPMS) and the Juno Microwave Radiometer (MWR). See the main text for references.

Atmospheric processes are also efficient at removing $CH_4$. Photochemistry in both atmospheres is initiated by solar ultraviolet (UV) radiation that dissociates and ionizes both $N_2$ and $CH_4$, leading to the production of complex hydrocarbon species and haze (Luspay-Kuti et al. 2017; Mandt et al. 2021). At Pluto and Triton, the photochemical loss rate for $CH_4$ is ~10× greater than the loss rate for $N_2$, leading to a preferential loss of carbon compared to nitrogen (Krasnopolsky & Cruikshank 1995; Luspay-Kuti et al. 2017). Furthermore, atmospheric escape will favor methane loss over nitrogen because methane is lighter. The current escape rate of methane $CH_4$ observed by New Horizons is 2 orders of magnitude greater than the $N_2$ loss rate (Gladstone et al. 2016). Therefore, the removal of methane through atmospheric processes could reasonably contribute to the carbon depletion for both Pluto and Triton.

It is likely that both of these processes have been effective at removing $CH_4$ from both Pluto and Triton. The observed abundances of $CH_4$ and $NH_3$ relative to water are very similar in comets. If both were partially converted to $CO_2$ and $N_2$ in the interior, the bulk of the $N_2$ observed would have been derived from $NH_3$, while the remaining $CH_4$ would have been primordial in origin. Trapping of $CO_2$ in clathrates in a subsurface ocean (Kamata et al. 2019) followed by additional $CH_4$ loss through atmospheric processes could explain the carbon depletion observed in N/C and O/C at Pluto and Triton.

### 3.3. Water and Building Block Analogs

Glein & Waite (2018) proposed based on the $N_2/H_2O$ ratio measured for comet 67P/C-G that the $N_2$ observed at Pluto was primordial and that the high $N_2/CO$ ratio was the result of loss processes for CO. We evaluate here the water abundance of a wider range of building block analogs to determine how the total nitrogen from the solid building blocks, whether in the form of $N_2$, $NH_3$, and HCN ices or from organics in chondritic materials compares to the water that has been observed for these building blocks. We do this by comparing the N/O and O/H ratios for Pluto and Triton with the analogs for the solid building blocks and for solar and Jupiter values in Figure 2. We can see from this figure that the observed N/O and O/H are essentially the same as that of cometary ice. This suggests that the observed nitrogen likely came primarily from $NH_3$ ices with some contribution of $N_2$ and possibly organics.

Although Triton's error bars almost overlap with Pluto's, it is notable that Triton's N/O is larger than Pluto's by almost 1 order of magnitude. This supports the previous observation based on the carbon depletion and models for aqueous chemistry that Triton's primordial $NH_3$ and organics may have been more efficiently converted to $N_2$ than Pluto's.

### 3.4. Nitrogen Isotopes

The most valuable constraint for the origin of nitrogen is the current nitrogen isotope ratio, $^{14}N/^{15}N$, in $N_2$ at Pluto and Triton. Although the measurement, if made, would not be the primordial value, by constraining the effects of the evolution of the atmosphere it can be used to constrain the origin of nitrogen (e.g., Mandt et al. 2014 for Titan). The primordial nitrogen isotope ratio when Pluto and Triton formed would have a ratio of ∼450 if the nitrogen originated as $N_2$, while an origin of $NH_3$ would have a primordial ratio of ∼130. Organics would have an intermediate ratio in the range of ∼250 (see Mandt et al. 2017 and references therein). The escape of particles from the top of the atmosphere will preferentially remove the light isotope, causing the isotope ratio to become lower over time. On the other hand, photochemistry can remove more of the heavy isotope, making the ratio larger over time. Mandt et al. (2017) found that the ALMA upper limit for the $HC^{15}N$ abundance (Lellouch et al. 2017) is not sufficient to constrain the source of nitrogen in Pluto and future observations of the nitrogen isotopes in $N_2$ are needed.

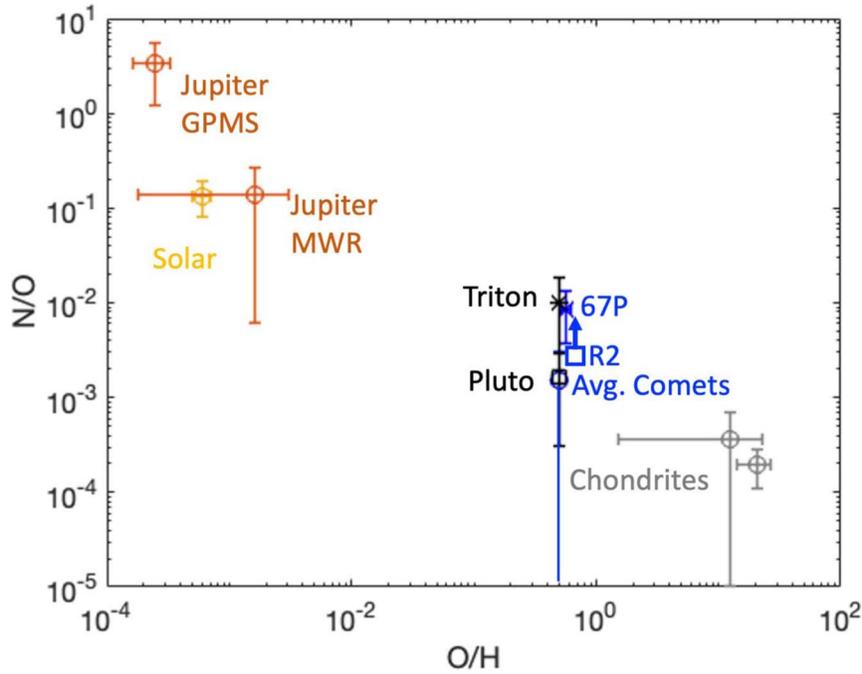

**Figure 2.** Elemental ratios providing information on the relative abundance of nitrogen compared to water for Pluto and Triton. This is compared to the solar abundances, abundances observed in Jupiter's atmosphere, and analogs for solid materials in the PSN.

### 3.5. Argon Depletion

The final observation that is relevant to understanding the formation conditions of the building blocks for Pluto and Triton is the upper limit for argon observed by New Horizons (Steffl et al. 2020). Because this observation is for 400 km from the surface, atmospheric models are required to determine the upper limit for the surface value of argon. The main uncertainty is what models assume for the eddy diffusion coefficient, which is a parameter used in one-dimensional models to approximate the dynamical mixing in an atmosphere. This parameter is notoriously difficult to constrain as demonstrated by the variety of profiles derived for Titan (see Figure 5 of Mandt et al. 2022). For Pluto, the published profiles vary by 3 orders of magnitude with major implications for the argon upper limit at the surface. As noted in Steffl et al. (2020), the low values for eddy diffusion assumed by Wong et al. (2017) and Young et al. (2018) allow for upper limits for argon that are meaningless for origins studies. However, in Luspay-Kuti et al. (2017) a high eddy diffusion coefficient, similar to the value derived by Gladstone et al. (2016), provided the best fit to the methane and hydrocarbon profiles. The authors were not able to reproduce the methane profiles published by Wong et al. (2017) using low eddy diffusion coefficients without requiring a high flux of methane from the surface ices. Further work is needed within the modeling community to resolve these issues. If the high eddy diffusion values are valid, the upper limit for argon on the surface can be used to estimate the lower limit for N/Ar illustrated in Figure 3 where it is compared to solar values and analogs for solid building blocks.

Assuming the ratio has not evolved over time, this lower limit would rule out an origin of the nitrogen being delivered primarily as $N_2$, because the ratio would have been solar in composition (Lodders 2021). As discussed earlier, there are processes that would change elemental ratios over time. Changing N/Ar requires preferential removal of either nitrogen or argon. Nitrogen can be preferentially removed by photochemistry and escape, and could decrease the N/Ar ratio significantly over time from a higher primordial value to whatever the value is today. If this is the only process changing N/Ar over time it would rule out a solar primordial ratio.

Processes that preferentially remove argon are limited. Mousis et al. (2013) explored the removal of noble gases from the atmosphere by trapping them in clathrates on the surface. However, $N_2$ and Ar have similar stability in clathrates, so it is unlikely that this would change the N/Ar enough to move a solar ratio to a value higher than the lower limit shown in Figure 3. Formation of Ar ice deposits on the surface could also remove Ar from the atmosphere, but again would be at a similar rate to $N_2$, which condenses at similar temperatures. Therefore, this supersolar lower limit for N/Ar is likely primordial and reflective of the solid building block composition for Pluto, with similar implications to Triton.

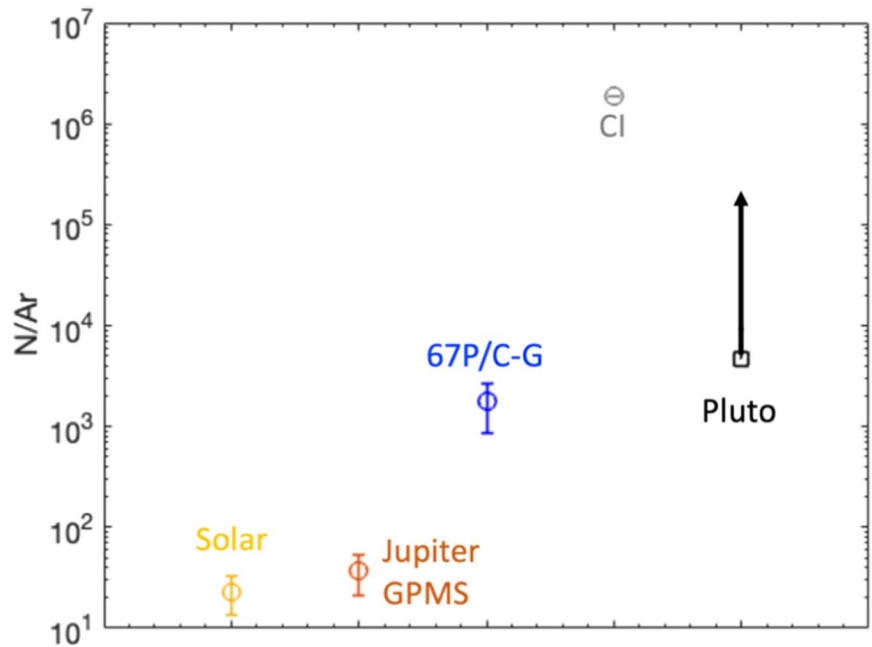

**Figure 3.** The N/Ar lower limit for Pluto compared to solar values, Jupiter, and 67P/C-G and CI chondrites, which serve as analogs for ices and organics in the solid building blocks that may have formed Pluto.

## 4. Summary and Conclusions

Previous work evaluated the abundance of $N_2$ at Pluto relative to the estimated bulk abundance of water and concluded that Pluto's $N_2$ was delivered to the building blocks in this form rather than as $NH_3$ or organics (Glein & Waite 2018). However, a more complex picture emerges when considering a more comprehensive set of available measurements and upper limits not only at Pluto but also at Triton and for comets. Furthermore, using elemental ratios rather than molecular abundances eliminates many processes that are difficult to constrain over geologic timescales.

Comets contain much more $NH_3$ than $N_2$, greatly increasing the bulk amount of nitrogen relative to oxygen (N/O) and carbon (N/C) in comets than when only considering $N_2/H_2O$. By considering all species, we note three observations that point to an origin of $NH_3$ for the $N_2$ observed on the surface and in the atmosphere of both. First is that the observed volatile abundances are deficient in carbon relative to any analog for solid building blocks, even if the solid building blocks formed in cold enough conditions to efficiently trap $N_2$. Although the removal of CO and $CO_2$ could explain the high N/C ratio, this cannot explain the extreme enrichment in O/C. This can only be explained by the loss of $CH_4$, which likely occurred as a result of a combination of aqueous chemistry producing $CO_2$ that was later trapped in clathrates, and lost from the atmosphere through photochemistry and escape. Comparison of N/O and O/H for Pluto and Triton with potential primordial values suggests that the current $N/H_2O$ is in agreement with the bulk nitrogen to water abundance in comets and chondrites, which are reasonable analogs for their building blocks. Finally, if the N/Ar lower limit is properly constrained using a high value for eddy diffusion for Pluto's atmosphere, the value derived also agrees with the building blocks having a cometary or chondritic composition, which would mean that Pluto's bulk nitrogen originated as $NH_3$ ices, with possible contributions from organics. More work is needed by the modeling community to resolve disagreements between models for eddy diffusion values not only for Pluto but also for Titan.

The observed volatile composition has further implications for the history of aqueous chemistry at Pluto and Triton. The lower abundance of $CH_4$ at Triton compared to Pluto, combined with the detection of $CO_2$ at Triton but not at Pluto suggests that aqueous chemistry in the interior of Triton was more effective at converting $CH_4$ to $CO_2$ and $NH_3$ to $N_2$. This would indicate that tidal heating in Triton has allowed liquid phase chemistry in the interior to last longer than at Pluto. These results have applications to other large KBOs like Eris, where $N_2$ ices have been detected on the surface. The presence of $N_2$ on the surface of a KBO could indicate sufficient internal heating to result in aqueous chemistry that converts $CH_4$ to $CO_2$ and $NH_3$ to $N_2$.

## Acknowledgments

K.E.M. and A.L.K. acknowledge support from NASA NFDAP grant 80NSSC18K1233.

## ORCID iDs

Kathleen Mandt https://orcid.org/0000-0001-8397-3315
Adrienn Luspay-Kuti https://orcid.org/0000-0002-7744-246X
Olivier Mousis https://orcid.org/0000-0001-5323-6453
Sarah E. Anderson https://orcid.org/0000-0002-9189-581X